\documentclass[aps,prl,twocolumn,floatfix,footinbib,notitlepage,superscriptaddress,groupaddress,showpacs]{revtex4-1}
\usepackage{times,graphicx,graphics,xcolor,amsfonts,amssymb,amsbsy,amsthm,amsmath,dsfont,hyperref,bm,bbm,color}
\hypersetup{
    colorlinks,
    linkcolor={blue},
    citecolor={blue},
    urlcolor= {blue}
}
\newcommand{\ket}[1]{\left|{#1}\right\rangle}

\newcommand{\ignore}[1]{}

\let\oldsqrt\sqrt
\def\sqrt{\mathpalette\DHLhksqrt}
\def\DHLhksqrt#1#2{%
\setbox0=\hbox{$#1\oldsqrt{#2\,}$}\dimen0=\ht0
\advance\dimen0-0.2\ht0
\setbox2=\hbox{\vrule height\ht0 depth -\dimen0}%
{\box0\lower0.4pt\box2}}

\DeclareFontFamily{OT1}{pzc}{}
\DeclareFontShape{OT1}{pzc}{m}{it}%
              {<-> s * [1.25] pzcmi7t}{}
\DeclareMathAlphabet{\mathpzc}{OT1}{pzc}%
                                 {m}{it}

\begin{document}

\title{Quantum gate description for induced coherence without induced emission and related phenomena}

\author{Sahar Alipour}
\email{salipour@ipm.ir}
\affiliation{School of Nano Science, Institute for Research in Fundamental Sciences (IPM), Tehran 19538, Iran}
\author{Mario Krenn}
\email{mario.krenn@univie.ac.at}
\affiliation{Vienna Center for Quantum Science \& Technology (VCQ), Faculty of Physics, University of Vienna, Boltzmanngasse 5, 1090 Vienna, Austria}
\affiliation{Institute for Quantum Optics and Quantum Information (IQOQI), Austrian Academy of Sciences, Boltzmanngasse 3, 1090 Vienna, Austria}
%\author{A. T. Rezakhani}
%\affiliation{Department of Physics, Sharif University of Technology, Tehran 14588, Iran}
\author{Anton Zeilinger}
\affiliation{Vienna Center for Quantum Science \& Technology (VCQ), Faculty of Physics, University of Vienna, Boltzmanngasse 5, 1090 Vienna, Austria}
\affiliation{Institute for Quantum Optics and Quantum Information (IQOQI), Austrian Academy of Sciences, Boltzmanngasse 3, 1090 Vienna, Austria}
%%%%%%%%%%%%%%%%%%%%%%%%%%%%%%%%%%%%%%%%%%%%%%%%%%%%%%%%%%%%%%%%

\begin{abstract}
We introduce unitary quantum gates for photon pair creation in spontaneous parametric down-conversion nonlinear crystals (NLs) and for photon path alignment. These are the two key ingredients for the method of \textit{induced coherence without induced emission} and many ensuing variations thereof. The difficulty in doing so stems from an apparent mixing of the mode picture (such as the polarization of photons) and the Fock picture (such as the existence of the photons). We illustrate utility of these gates by obtaining quantum circuits for the experimental setups of the frustrated generation of photon pairs, identification of a point-like object with undetected photons, and creation of a Bell state. We also introduce an effective nonunitary description for the action of NLs in experiments where all the NLs are pumped coherently. As an example, by using this simplifying picture, we show how NLs can be used to create superposition of given quantum states in a modular fashion.
\end{abstract}

\pacs{42.50.-p, 
%quantum optics
03.67.-a, 
%QI
%03.65.Wj,
%state reconstruction, tomography
34.80.Pa, 
%coherence ad correlation
%03.67.Lx, 
%Quantum computation architectures and implementations
%42.50.Ex
%Optical implementations of quantum information processing and transfer
42.30.Wb
%Image reconstruction, tomography
}
\date{\today}
\maketitle

%%%%%%%%%%%%%%%%%%%%%%%%%%%%%%%%%%%%%%%%%%%%%%%%%%%%%%%%%%%%%%%%
\textit{Introduction.---}Recently, interest in the ``induced coherence without induced emission" \cite{ZWM:ICWIE1, ZWM:ICWIE2} has been revived due to its various renewed applications, e.g., in quantum imaging \cite{Lemos, Lahiri:imaging,Controlling}, generation of entanglement in multipartite and high-dimensional systems \cite{Krenn:path identity}, connections to graphs and computational complexity \cite{Gu}, quantum spectroscopy \cite{Kulik}, investigation of the complementarity relation \cite{Menzel}, novel methods to quantify quantum correlations \cite{armin}, and recently even in superconducting microwave resonators \cite{lahteenmaki2016coherence}. 
Central in the original experiment was overlapping two paths; one output path from each of a pair of nonlinear spontaneous parametric down-conversion crystals (NLs). This overlapping or ``path alignment" is responsible for inducing coherence in the output state of this experiment. 

With the advent of quantum computation and extensive efforts for its optical realizations \cite{KLM,OBrien,Ralf-Pryde,Ralph,exp-one-way}, and noting the importance of the coherence generation, it is imperative to understand the action of the path alignment and correlated pair creation in NLs within the context of quantum computation.
To this end, one needs to  attribute quantum gates to these effects.
Although quantum circuits for several specific quantum-optical experiments containing NLs have already been proposed \cite{Pryde-Fredkin,Ghalaii}, a systematic approach for designing such quantum circuits is still lacking, mainly due to mixing of polarization and path degrees of freedom of photons which are created on-the-fly.

Here we solve this problem systematically. 
First, by defining photonic qutrits based on polarization and photon numbers in each path, we lay out a reduced unitary description for coherent creation of photon pairs, assuming weak nonlinearity or first-order approximation in the NL Hamiltonian. Besides, we introduce a unitary picture which describes path alignment. These gates enable one to attribute systematically unitary quantum circuits \cite{Nielsen-Chuang} to quantum-optical experiments containing NLs and path alignment and helps to bridge the gap between quantum-optical experiments and quantum computation. We employ these gates for presenting quantum unitary circuits for three important quantum-optical experiments.

Second, we introduce an \textit{effective} picture for creation of photon pairs. This picture relies on the fact that vacuum states cannot be detected in experiments. In addition, in contrast to the unitary approach, in this picture experiments can be described with photonic qubits using only the polarization degrees of freedom. An important feature of this effective picture are is nonlinearity of the action of an NL. Nevertheless, in special cases one can attribute a linear (yet nonunitary) operator to an NL.  As an application, using this effective picture, we show that how to modularly create a superposition of a set of given quantum states. The features of this effective picture may allow one to employ it for realization of tasks which may seem impossible otherwise. 

%%%%%%%%%%%%%%%%%%%%%%%%%%%%%%%%%%%%%%%%%%%%%%%%%%%%%%%%%%%%%%%%
\textit{Unitary description of an NL.---}The Hamiltonian of an NL is $H_{\textsc{nl}}=g a_p a^{\dag}_s a^{\dag}_i+g^{\ast}a^{\dag}_p a_s a_i$, where $g$ denotes the down-conversion factor, which is often assumed $\ll 1$, and indices $p$, $s$, and $i$ denote distinguishable pump, signal, and idler photons, respectively \cite{Walborn}. For simplicity, we assume throughout this paper that the creation operators $a^{\dag}_p$, $a^{\dag}_s$, and $ a^{\dag}_i$ create photons, using the type-I phase matching, only with horizontal polarization \cite{Kwiat-Steinberg}. Thus the action of an NL is given by
\begin{equation}
U_{\textsc{nl}}=e^{-i H_{\textsc{nl}}}=\openone - i g a_p a^{\dag}_s a^{\dag}_i- ig^{\ast} a^{\dag}_p a_s a_i +O(|g|^2).
\label{U-NL}
\end{equation}

We assume that the pump photon is used only to activate NL and that the initial state is $|\alpha \rangle_p |0\rangle_s |0\rangle_i$, where $|\alpha\rangle$ is a coherent state. 
In addition, suppose that the laser is sufficiently weak, i.e., $|\alpha g|\ll 1$, and hence we can keep terms only up to the first order of $|\alpha g|$. These assumptions lead to the following conclusions: (i) in a multi-NL setup which is pumped \textit{coherently} \cite{ZWM:ICWIE2}, existence of a signal or idler photon in any step of the experiment implies that one of the NLs in some previous step has fired, which in turn means that this state is already of $O(|g\alpha|)$. In this case, the application of the next $U_{\textsc{nl}}$ on this state is given by the identity operator. Equivalently, there is at most one pair of photons in the whole experiment. (ii)
 The state of the pump photon does not change during the experiment. Thus it is not necessary to include pump in the matrix representation of the associated unitary gate for an NL. 
In contrast to pump photons, since we are interested in signal and idler photons, their polarization state may be manipulated by using suitable waveplates. Bearing in mind that there exists at most one created pair in the setup, one can describe the quantum state of each path as a \textit{qutrit} in terms of three possible quantum states: $\ket{0}$ when there is no photon in the path, $|H \rangle$ when there is one horizontally-polarized photon, and $\ket{V}$ when there is one vertically-polarized photon in the path.

\begin{figure}[tp]
\includegraphics[scale=0.35]{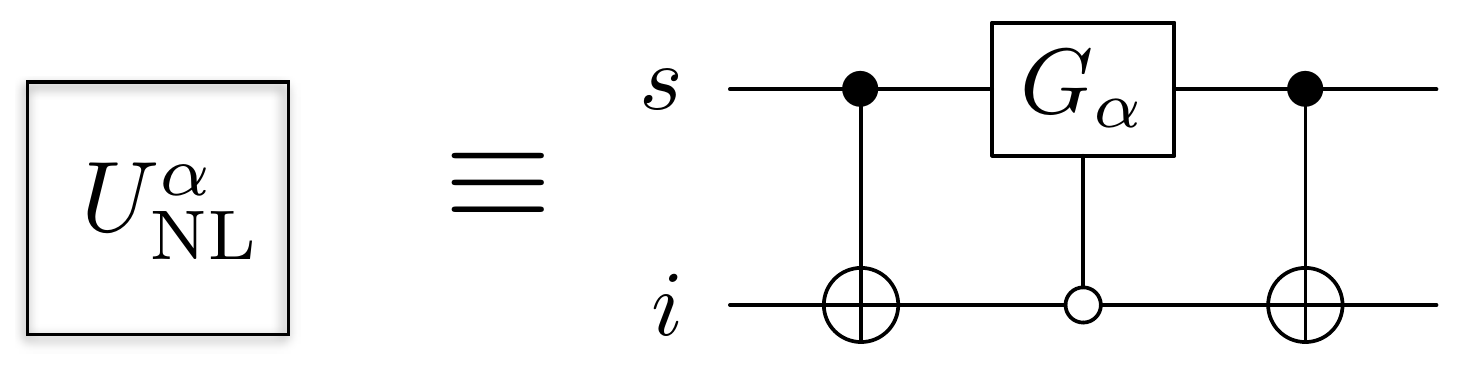}
\caption{Unitary gate for an NL. Here a filled circle implies that the corresponding $\textsc{not}$ gate (which transforms $|H\rangle$ and $|0\rangle$ to each other) applies on the target photon when the state of the control photon is $\ket{H}$; otherwise, it acts as an identity.}
\label{NL-alpha-gate}
\end{figure}

It is evident from Eq.~\eqref{U-NL} and the above assumptions that the unitary transformation induced by an NL, in the space spanned by $\{ |kl\rangle\}_{k,l=0,H,V}$, can be described with the quantum gate
%%%%%%%%%%%%%%%
\begin{align}
U_{\textsc{nl}}^{\alpha} |H H\rangle&= |H H\rangle - i g^{\ast} \alpha^{\ast} |00 \rangle, \nonumber\\
U_{\textsc{nl}}^{\alpha} |00 \rangle&=|00 \rangle - i g \alpha |H H\rangle,
\end{align}
%%%%%%%%%%%%%%%
and it applies on the seven other basis states as identity. Figure.~\ref{NL-alpha-gate} shows the quantum circuit for the action of  $U_{\textsc{nl}}^{\alpha}$,
%---------------------------------------------------------------------------------------
%---------------------------------------------------------------------------------------
where the generalized-$\textsc{cnot}$ ($\textsc{g-cnot}$) gate is defined through $\textsc{g-cnot}\ket{H 0}=\ket{H H}$ and $\textsc{g-cnot}\ket{H H}=\ket{H 0}$ (and identity for the rest of states). In the qutrit basis we have the matrix representations
%%%%%%%%%%%%%%%%
\begin{align}
\textsc{g-cnot}=\left(\begin{smallmatrix}
\openone_{3\times 3} &  &  & \\
 &  0 & 1 &  \\
 &  1 & 0 &  \\
 &     &    & \openone_{4\times 4} \\
\end{smallmatrix}
\right),~~~
G_{\alpha}=\left(\begin{smallmatrix}
1 & -ig^{\ast}\alpha^{\ast}  & 0\\
-ig \alpha& 1 & 0 \\
0 & 0 & 1
\end{smallmatrix}
\right).
\end{align}
%%%%%%%%%%%%%%%%%
%%%%%%%%%%%%%%%%%%%%%%%%%%%%%%%%%%%%%%%%%%%%%%%%%%%%%%%%%%%%%%%%
\textit{Quantum gate for path alignment.---}When one of the output paths ($\ell_1$) of an NL  is aligned with the corresponding input path ($\ell_2$) of another NL, the effect of this alignment is to transfer photons from $\ell_1$ to $\ell_2$. This action can be described with a two-qutrit \textsc{swap} gate 
which acts nontrivially only on the subspace defined by $\{|H 0\rangle,|0 H\rangle,|V 0\rangle,|0 V\rangle\}_{\ell_{1}\ell_{2}}$---and it acts as identity on the other possible states (although here they do not occur).
%---------------------------------------------------------------------------------------
\begin{figure}[tp]
\includegraphics[scale=0.3]{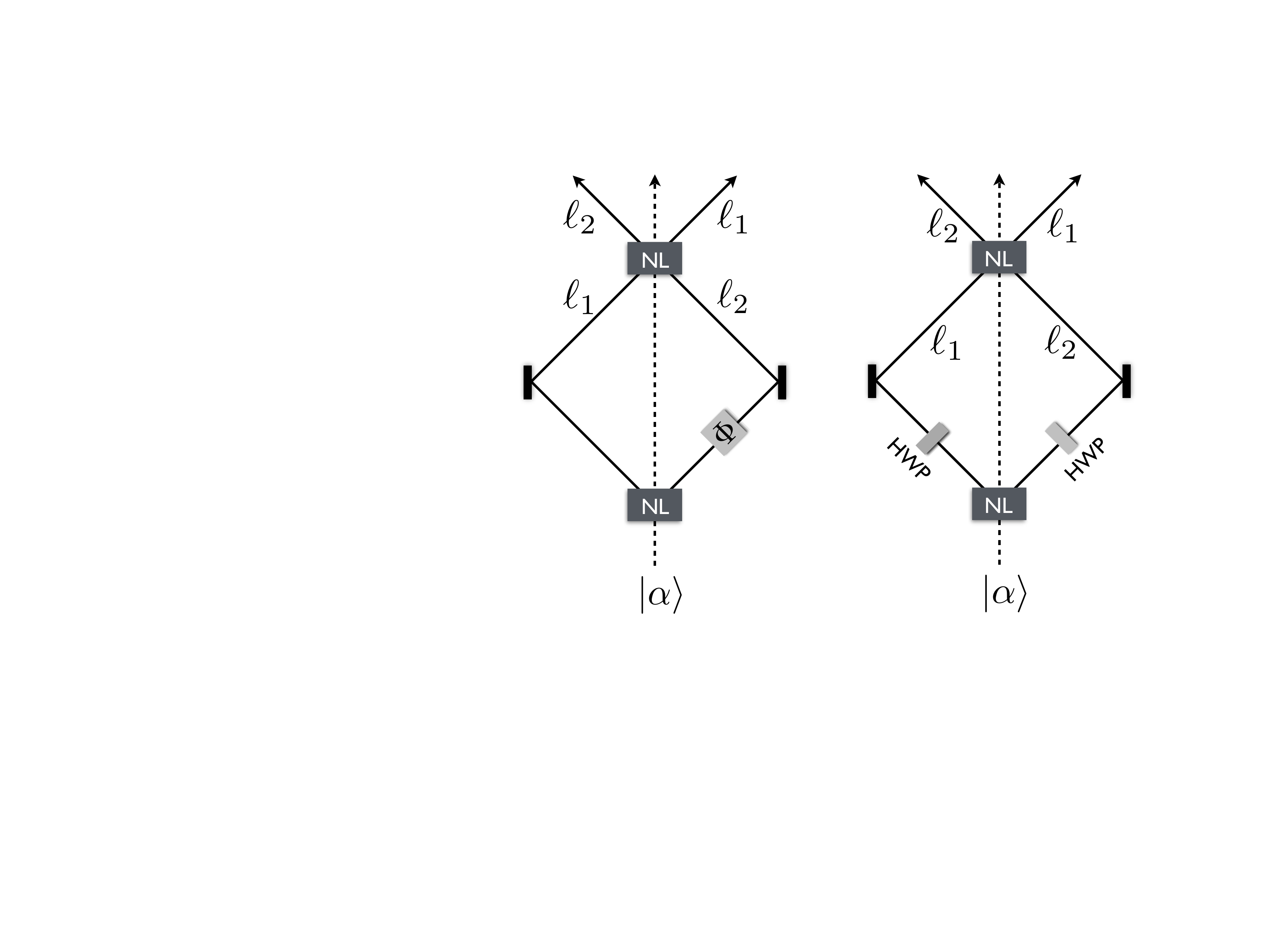}
\caption{Setups for (left) frustrated generation of photon pairs and (right) generation of a maximally-entangled pair.}
\label{Bell-state-setup}
\end{figure}
%---------------------------------------------------------------------------------------

In the following, we illustrate how to employ these quantum gates in three important examples. To help clarify the action of path alignment in these experiments, we draw separate signal and idler paths for each NL even if this might seem redundant, and apply path alignment gate where it is needed. 
%%%%%%%%%%%%%%%%%%%%%%%%%%%%%%%%%%%%%%%%%%%%%%%%%%%%%%%%%%%%%%%%

\begin{figure}[bp]
\includegraphics[scale=0.45]{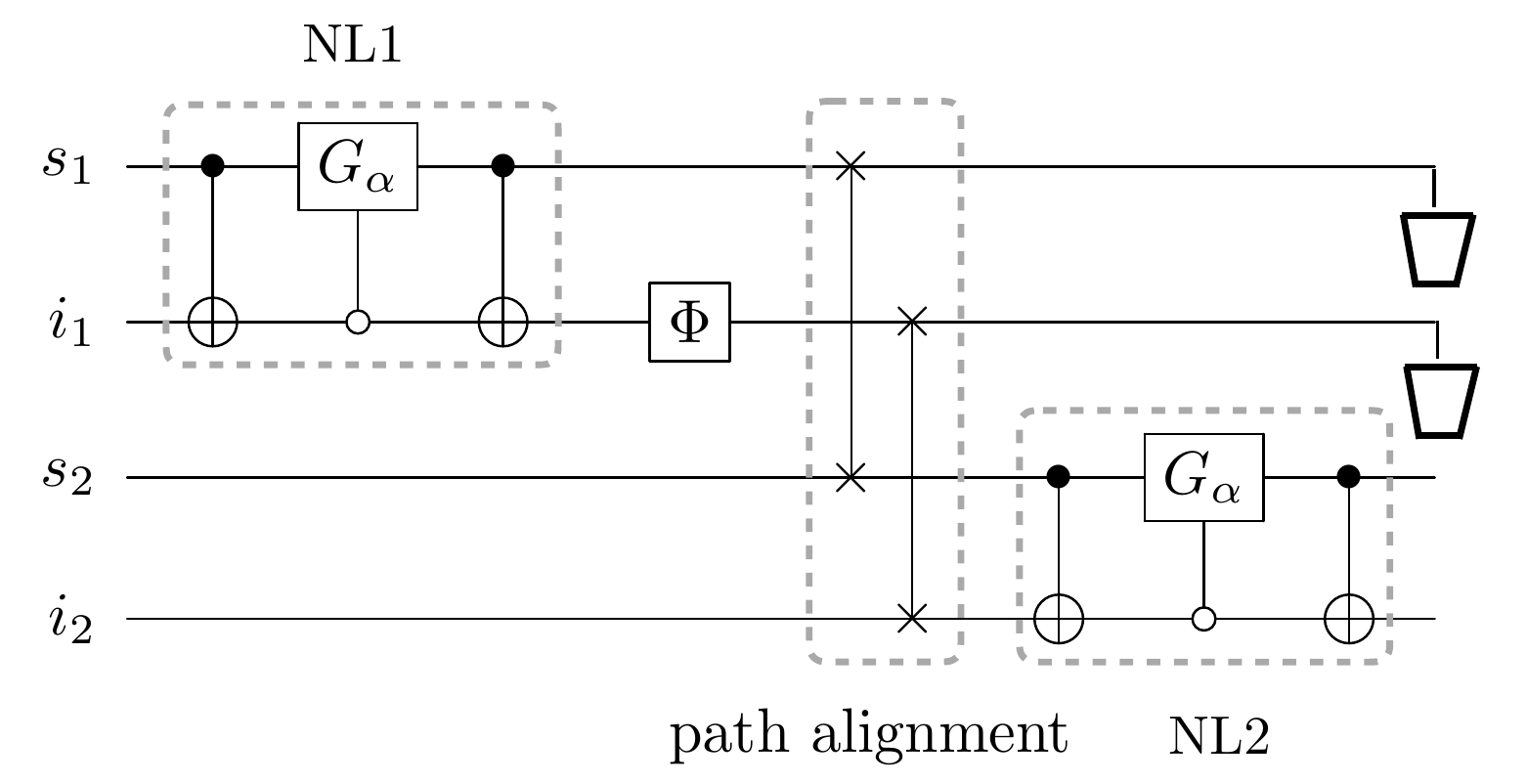}
\caption{
Quantum circuit for frustrated down-conversion.
}
\label{frustrated-mandel-alpha}
\end{figure}

\textit{Example I: Quantum circuit for frustrated generation of photon pairs.---}Using the introduced elements of the quantum circuits, $U_{\textsc{nl}}^{\alpha}$ and $\textsc{swap}$, one can see that the frustrated down-conversion \cite{frustrated-Mandel}---Fig.~\ref{Bell-state-setup} (left)---can be described with the quantum circuit depicted in Fig.~\ref{frustrated-mandel-alpha}. 
%---------------------------------------------------------------------------------------
%--------------------------------------------------------------------------------------- 
Since the state of the pump throughout the experiment remains as the initial pump state $|\alpha \rangle$, we can remove it from the state representation of the photons. 
Assuming $|\psi_0\rangle= |00\rangle_{s_1 i_1} |00\rangle_{s_2 i_2}$ as the initial state for the quantum circuit of Fig.~\ref{frustrated-mandel-alpha} yields the following states in the corresponding steps of the experiment:
%%%%%%%%%%%%%%%%%%%%%%%%%%%
\begin{align}  
|\psi_1\rangle&:=U_\textsc{nl1}^{\alpha} |\psi_0\rangle=(|00\rangle -ig \alpha |H H\rangle)_{s_1 i_1} |00\rangle_{s_2 i_2}
\label{psi-1},\\
|\psi_2\rangle&:= U_{\Phi} |\psi_1\rangle=(|00\rangle - i e^{-i\Phi} g \alpha |H H\rangle)_{s_1 i_1} |00\rangle_{s_2 i_2},\nonumber\\
|\psi_3\rangle&:=(\textsc{swap}_{s_1 s_2} \textsc{swap}_{i_1 i_2})|\psi_2\rangle =|00\rangle_{s_1 i_1} \nonumber\\
&\quad\times(|00\rangle-i e^{-i\Phi}g\alpha |H H\rangle)_{s_2 i_2}, \nonumber\\
|\psi_4\rangle&:=U_\textsc{nl2}^{\alpha}|\psi_3\rangle
=|\psi_0\rangle -i g \alpha (1+e^{-i\Phi} ) |00\rangle |H H\rangle.
\label{frus-mandel}
\end{align}
It is seen that by choosing $\Phi=\pi$ in the final state $|\psi_4\rangle$ one can reach the initial state $|\psi_0\rangle$. That is no photon can be detected in either of the signal or idler paths. 

%%%%%%%%%%%%%%%%%%%%%%%%%%%%%%%%%%%%%%%%%%%%%%%%%%%%%%%%%%%%%%%%
\textit{Example II: Quantum circuit for creation of a maximally-entangled state.---}A setup for creation of maximally-entangled state between the polarizations of two photons is given in Fig.~\ref{Bell-state-setup} (right) \cite{kwiat1999ultrabright}, and its circuit-based description is similar to what depicted in Fig.~\ref{frustrated-mandel-alpha} with this difference that $\Phi$ should be replaced with two half-waveplates (\textsc{hwp}s) in paths $s_1$ and $i_1$. The action of \textsc{hwp} is $\textsc{hwp}|H\rangle =|V\rangle$ and $\textsc{hwp}|V\rangle =|H\rangle$ (and obviously $\textsc{hwp}|0\rangle =|0\rangle$), which is the Pauli $X$ operator in the $\{|H\rangle, |V\rangle\}$ endocing. Following steps akin to those of Eq.~\eqref{frus-mandel}, it is straightforward to see that the final state will be $|00\rangle_{s_1i_1} |00\rangle_{s_2i_2}-ig\alpha |00\rangle _{s_1i_1}(|H H\rangle + |V V\rangle)_{s_2i_2}$, which means that the polarization state of the two photons of paths $s_2$ and $i_2$ are maximally-entangled. Note that the probability of generating this state is of $O(|g \alpha |^2)$.

\begin{figure}[tp]
\includegraphics[scale=0.34]{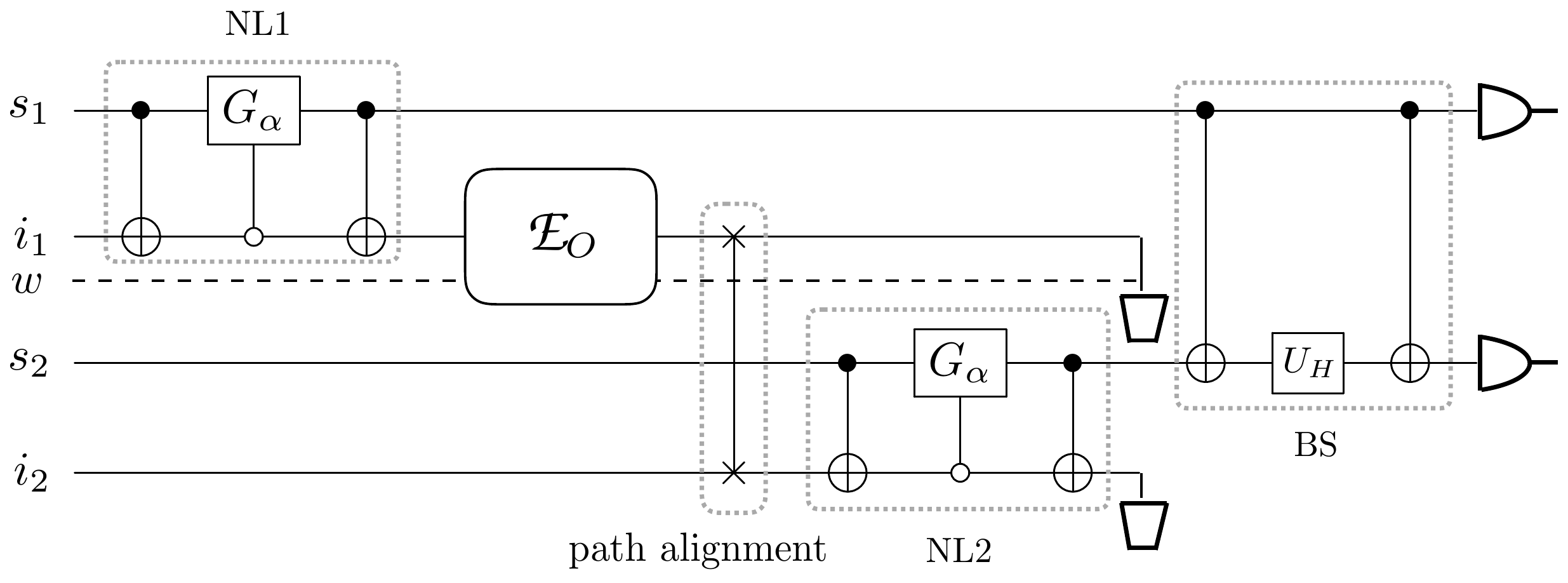}
\caption{Quantum circuit for identification of a point-like object with undetected photons. $\mathpzc{E}_O$ is the object, $U_H$ is the Hadamard gate, and BS denotes beam splitter.}
\label{q-imaging-corr}
\end{figure}

%%%%%%%%%%%%%%%%%%%%%%%%%%%%%%%%%%%%%%%%%%%%%%%%%%%%%%%%%%%%%%%%
\textit{Example III: Quantum circuit for identification of a point-like object with undetected photons.---}The quantum circuit proposed in Fig.~\ref{q-imaging-corr} can describe identification of a point-like object with unknown transmittance $T$ and phase shift $\gamma$ parameters such that $\mathpzc{E}_O\ket{1 0}_{iw}=Te^{i\gamma}\ket{10}_{iw}+\sqrt{1-T^2}\ket{01}_{iw}$ \cite{Ghalaii,note}. We remark that a real object can be approximated as a combination of point-like objects on a planar region \cite{Lahiri:imaging}. Hence this circuit can enable one to analyze the more sophisticated ``quantum imaging with undetected photons experiment." Details of the states of the system in each step is given in Ref.~\cite{supp}.
%---------------------------------------------------------------------------------------
%---------------------------------------------------------------------------------------
%%%%%%%%%%%%%%%%%%%%%%%%%%%%%%%%%%%%%%%%%%%%%%%%%%%%%%%%%%%%%%%%

\textit{Effective picture for describing setups with NLs.---}Since vacuum cannot be detected in quantum-optical experiments, it seems easier to work in an \textit{effective} picture in which the vacuum states are neglected. We remind that all the assumptions and remarks we discussed after Eq.~\eqref{U-NL}---coherently pumping setups and retaining $O(|g\alpha|)$ terms---are applied here too. In addition, in contrast to the unitary approach, in this picture we use only the polarization degrees of freedom for state representation.  

Let us define the effective state $|\psi^{\mathrm{eff}}\rangle$ as
\begin{align}
|\psi \rangle=|0\cdots 0\rangle -ig\alpha|\psi^{\mathrm{eff}}\rangle,
\end{align}
in which $|\psi\rangle$ is the state of the system. The action of an NL placed on the signal and idler paths  $\ell_i \ell_j$ reads
%%%%%%%%%%%%%%%%%%
\begin{equation}
(\Lambda^{\alpha}_{\textsc{nl}})_{\ell_i \ell_j} |\psi^{\mathrm{eff}}_{\mathrm{in}}\rangle= |\psi^{\mathrm{eff}}_{\mathrm{in}}\rangle+|H H\rangle_{\ell_i \ell_j},
\label{eff-NL}
\end{equation}
%%%%%%%%%%%%%%%%%%
in which $|\psi^{\mathrm{eff}}_{\mathrm{in}}\rangle$ is an arbitrary input state. In our notation, the state of the system on the unwritten paths is vacuum; e.g., by $|H H\rangle_{\ell_i \ell_j}$ we mean $|H H\rangle_{\ell_i \ell_j}|0\rangle^{{}^{\otimes N-2}}_{{}_{\overline{\ell_i \ell_j}} }$ in which $\overline{\ell_i \ell_j}$ indicates all paths except $\ell_i$ and $\ell_j$, and $N$ is the total number of paths. 

We note that $\Lambda^{\alpha}_{\textsc{nl}}$ is nonunitary. Besides, $\Lambda^{\alpha}_{\textsc{nl}}$ is \textit{nonlinear} in the sense that $\Lambda^{\alpha}_{\textsc{nl}} (c_1 |\psi^{\mathrm{eff}}_1\rangle+c_2 |\psi^{\mathrm{eff}}_2\rangle) \neq c_1 \Lambda^{\alpha}_{\textsc{nl}} |\psi^{\mathrm{eff}}_1\rangle + c_2 \Lambda^{\alpha}_{\textsc{nl} }|\psi^{\mathrm{eff}}_2\rangle$. Due to the nonlinearity property, $\Lambda^{\alpha}_{\textsc{nl}}$ cannot be described as a two-photon gate and does not admit a matrix representation in the ordinary sense; rather, it can be considered as a \textit{superposer} (which adds $|H H\rangle_{\ell_i \ell_j}$ to input states).  

\textit{Remark.---}However, it is possible to obtain a matrix representation for $\Lambda_{\textsc{nl}}^{\alpha}$ in some restricted sense.
\cite{supp}. 
Here we explain the simplified
case where the system has only two paths. In this case, the nonlinearity of $\Lambda^{\alpha}_{\textsc{nl}}$ can be bypassed by using an auxiliary extra dimension in an extended vector space. The effective state of the system in the extended space is represented as 
$\left(\begin{smallmatrix}
1 \\
  |\psi^{\mathrm{eff}}\rangle \\
\end{smallmatrix}\right)$, 
in which the first entry $1$ is the auxiliary element. With this, the effective operation of NL is given by the translation matrix 
\begin{align}
\Lambda_\textsc{nl}^{\alpha}=\left(\begin{smallmatrix}
1 & 0 &  \\
1 & 1 &  \\
   &    &  \openone_{3\times3}\\
\end{smallmatrix}
\right),
\end{align}
where $\{|a\rangle, |HH\rangle, |HV\rangle, |VH\rangle, |VV\rangle \}$ is used as the basis, with $|a\rangle=\begin{pmatrix}1 & 0 & 0 & 0 & 0\end{pmatrix}^{T}$ representing the vacuum (because $|\mathrm{vac}^{\mathrm{eff}}\rangle=\begin{pmatrix} 0 & 0 & 0 & 0\end{pmatrix}^T$). 
The matrix representation for $\Lambda_\textsc{nl}^{\alpha}$ can be generalized in systems with more than two paths  using its representation in a direct-sum vector space---see Ref.~\cite{supp}. 
%---------------------------------------------------------------------------------------
\begin{figure}[tp]
\includegraphics[scale=0.33]{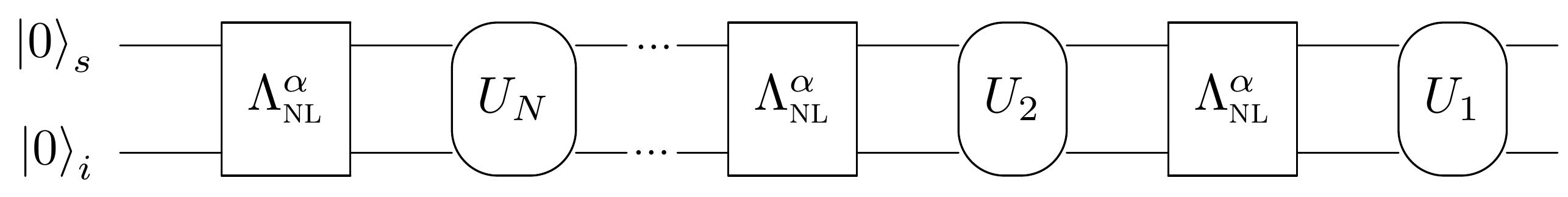}
\caption{
Quantum circuit for creation of uniform superposition.}
\label{superposition}
\end{figure}
%---------------------------------------------------------------------------------------

%%%%%%%%%%%%%%%%%%%%%%%%%%%%%%%%%%%%%%%%%%%%%%%%%%%%%%%%%%%%%%%%

\textit{An application of the effective picture: Modular creation of uniform superposition.---}Using the effective picture, one can describe how to create two-photon superposed polarization states \textit{modularly} by using a setup consisting of NLs. A high-dimensional generalization of this method will result in the generation of high-dimensional entangled states proposed in Ref. \cite{Krenn:path identity}. 

Let us assume that a set of two-photon states $\{\ket{\phi_i}\}_{i=1}^{N}$ is given and the objective is to create a uniformly superposed state as $\sum_{i=1}^{N} \ket{\phi_i}$ (omitting the normalization factor). To this end, we find a set of unitary operators $\{U^{(i)} \}_{i=1}^{N}$ such that $\ket{\phi_1} = U^{(1)} \ket{H H}$, ${U^{(1)}}^{-1}\ket{\phi_2} = U^{(2)} \ket{H H}$, ${U^{(2)}}^{-1} {U^{(1)}}^{-1}\ket{\phi_3} = U^{(3)} \ket{H H}$, \ldots, and ${U^{(N-1)}}^{-1}\ldots {U^{(2)}}^{-1} {U^{(1)}}^{-1} \ket{\phi_{N}}= U^{(N)} \ket{H H}$. 
Figure \ref{superposition} depictes the quantum circuit achieving the desired superposed state \cite{supp}.
An appealing feature of this superposition is that by removing the $i$th NL, the related state $\ket{\phi_i}$ will also be removed from the superposition without affecting other states.

It is straightforward to modify the setup of Fig.~\ref{superposition} to create similar superposition of given single-photon states. This can be achieved by applying single-photon unitary gates only on one of the paths (e.g., signal) and discarding the photon on the other (idler) path. To create a nonuniform superposition, it suffices to put the proper number of NLs successively (or equivalently put an NL in a cavity to use it multiply because of the reflection of photons). 

%%%%%%%%%%%%%%%%%%%%%%%%%%%%%%%%%%%%%%%%%%%%%%%%%%%%%%%%%%%%%%%%
\textit{Summary and outlook.---}We have obtained a unitary gate representation for photon pair creation with spontaneous parametric down-conversion in nonlinear crystals (NLs) which is suitable for describing experiments with weak pumping. We also have shown that path alignment can be unitarily described with a \textsc{swap} gate. We have used the introduced gates to obtain the equivalent quantum circuits for several quantum-optical experiments. Knowing the unitary gates for elements of experimental setups can be useful in designing quantum algorithms realizable by quantum optics. It can also be helpful in obtaining equivalent optical experimental setups for previously known quantum algorithms. Another application can be in designing computer programs for automated search for new quantum experiments \cite{Krenn-automated, Melnikov}.

Additionally, we have introduced an effective picture description for NLs which is more suitable for quantum-optical experiment. Using this picture it is simpler to see the application of NLs for creating modular superpositions of given quantum states. The effective picture may also be useful in constructing manybody quantum states with certain amount of entanglement \cite{Horodecki: ent, AKM-aklt}, such as graph states for measurement-based quantum computation \cite{MBQC,Nielsen:cluster}. 

A future question could involve investigation of the inherent nonlinearity of our effective description in the context of new quantum computation schemes. It will also be interesting to see how the effective description of processes can be generalized to multiphoton systems to allow employment of state-of-the-art photonic technology \cite{pan2012multiphoton, wang2016experimental}. Additionally, nonunitarity of the effective picture may also enable simulating open quantum dynamics more naturally within this picture.

\textit{Acknowledgements.---}S.A. thanks A. T. Rezakhani for useful discussions. S.A. also acknowledges the Austrian Academy of Sciences  (\"OAW) for the JESH fund and the IQOQI Vienna for its hospitality where this research was conducted. This work was supported by the \"OAW, the European Research Council (SIQS Grant No. 600645 EU-FP7-ICT), and the Austrian Science Fund (FWF) with SFB F40 (FOQUS). 

%%%%%%%%%%%%%%%%%%%%%%%%%%%%%%%%%%%%%%%%%%%%%%%%%%%%%%%%%%%%%%%%

%%%%%%%%%%%%%%%%%%%%%%%%%%%%%%%%%%%%%%%%%%%%%%%%%%%%%%%%%%%%%%%%
\begin{widetext}
\newpage

\appendix

%%%%%%%%%%%%%%%%%%%%%%%%%%%%%%%%%%%%%%%%%%%%%%%%%%%%%%%%%%%%%%%%
\section{Supplemental Materials}
\section{Appendix A: Unitary description of an NL when pump is a single photon}
 
As we have already explained in the main text,  one does not need to manipulate polarization of pump photons during the experiment, because pump is used only to activate NLs. Hence, if we assume the pump is a single photon, only two states $\ket{0}$ and $\ket{H}$ are sufficient to describe its state. With this assumption, the (truncated) Hilbert space for the possible input states of an NL with a single photon pump state, which occur in experimental setups with $g\ll 1$, in which all NLs pumped coherently, is spanned by
\begin{align}
\{ |000\rangle, |H00 \rangle, |0H0\rangle,  |0V0\rangle, |00H\rangle, |00V\rangle, |0HH\rangle,|0HV\rangle,|0VH\rangle,|0VV\rangle\}_{psi}.
\label{hnl}
\end{align}
Among these states the action of NL described by $U_{\textsc{nl}}^{\mathrm{P}}$ ($\textsc{p}$ indicates that pump is a single photon)) is nontrivial only on the subspace $\{|H00 \rangle, |0HH\rangle\}_{psi}$, and is given as the followings (for brevity, the subscripts are removed when it raises no ambiguity):
\begin{align}
U_{\textsc{nl}}^{\textsc{p}} |0HH\rangle&= |0HH\rangle - i g^{\ast} |H00 \rangle +O(|g|^2),\nonumber\\
U_{\textsc{nl}}^{\textsc{p}} |H00 \rangle&=| H00 \rangle - i g |0 HH \rangle +O(|g|^2),
\end{align}
and it applies on the other eight states of the subspace (\ref{hnl}) as identity. Using a method based on the Gray codes \cite{Nielsen-Chuang} the quantum circuit of Fig.~\ref{NL-P-1} for $U^{\mathrm{P}}_{\textsc{nl}}$ is obtained. In this figure, the three-photon $\textsc{g-cnot}$ gates with two control photons are natural generalization of two-photon $\textsc{g-cnot}$s. 
%---------------------------------------------------------------------------------------
\begin{figure}[bp]
\includegraphics[scale=0.3]{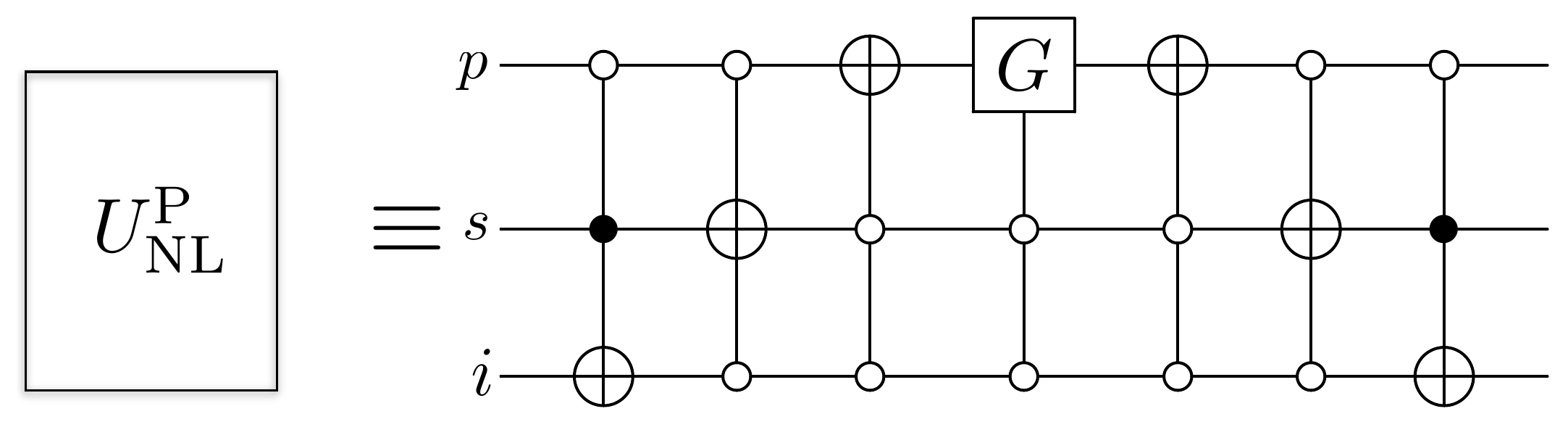}
\caption{Unitary gate for an NL when pump is a single photon. Filled circle denotes the corresponding $\textsc{not}$ gate applies on the target photon when the control photon is in $\ket{H}$, and unfilled circle denotes that the $\textsc{not}$ gate is applied on the target when the control photon is in $\ket{0}$; otherwise, it acts as an identity gate.
}
\label{NL-P-1}
\end{figure}
%---------------------------------------------------------------------------------------
The qutrit-gate $G$, in the computational basis which is defined in the main text, is given by
\begin{align}
G=\left(\begin{array}{ccc}
1 & -ig^{\ast} & 0 \\
-ig & 1 & 0 \\
0 & 0 & 1
\end{array}
\right). 
\label{G}
\end{align}

%%%%%%%%%%%%%%%%%%%%%%%%%%%%%%%%%%%%%%%%%%%%%%%%%%%%%%%%%%%%%%%%
\section{Appendix B: Quantum circuit for frustrated down-conversion when pump is a single photon}

The quantum circuit for frustrated down-conversion with a single photon pump (see Fig.~\ref{frustrated-mandel}) with quantum state $|H\rangle_{p_1}|0\rangle_{p_2}+|0\rangle_{p_1}|H\rangle_{p_2}$ is given in Fig.~\ref{frustrated-mandel}.
The effect of the circuit on the input $|\psi_0\rangle=|H00\rangle_{p_1 s_1 i_1} |000\rangle_{p_2 s_2 i_2}+|000\rangle_{p_1 s_1 i_1} |H00\rangle_{p_2 s_2 i_2}$ becomes
\begin{align}
|\psi_1\rangle&:= U_\textsc{nl1}^{\textsc{p}} |\psi_0\rangle=(|H00\rangle_{p_1 s_1 i_1} -ig |0HH\rangle_{p_1 s_1 i_1})|000\rangle_{p_2 s_2 i_2} +|000\rangle_{p_1 s_1 i_1} |H00\rangle_{p_2 s_2 i_2}\nonumber\\
|\psi_2\rangle&:= U_{\Phi} |\psi_0\rangle=(|H00\rangle_{p_1 s_1 i_1} - i e^{-i\Phi} g |0HH\rangle_{p_1 s_1 i_1})|000\rangle_{p_2 s_2 i_2} +|000\rangle_{p_1 s_1 i_1} |H00\rangle_{p_2 s_2 i_2}\nonumber\\
|\psi_3\rangle&:=(\textsc{swap}_{s_1 s_2} \textsc{swap}_{i_1 i_2})|\psi_1\rangle =|H00\rangle |000\rangle-i e^{-i\Phi}g |000\rangle |0HH\rangle+|000\rangle |H00\rangle\nonumber\\
|\psi_4\rangle&:=U_\textsc{nl2}^{\textsc{p}}|\psi_2\rangle=|H00\rangle |000\rangle-i e^{-i\Phi} g |000\rangle (|0HH\rangle-i g^{\ast}|H00\rangle) +|000\rangle (|H00\rangle-ig |0HH\rangle)\nonumber\\
&~=|\psi_0\rangle -i g (1+e^{-i\Phi} ) |000\rangle |0HH\rangle+O(|g|^{2}).
\end{align}
By choosing $\Phi=\pi$, it can be seen that the output state will become equal to the input state, i.e.,  $|\psi_4\rangle = |\psi_0\rangle$, which gives zero photon on both final paths $i_2$ and $s_2$. Besides, when the phase shifter is placed on path $s$ or even on the input pump path of one of the crystals, the same effect can be observed.
%---------------------------------------------------------------------------------------
\begin{figure}
\includegraphics[scale=0.31]{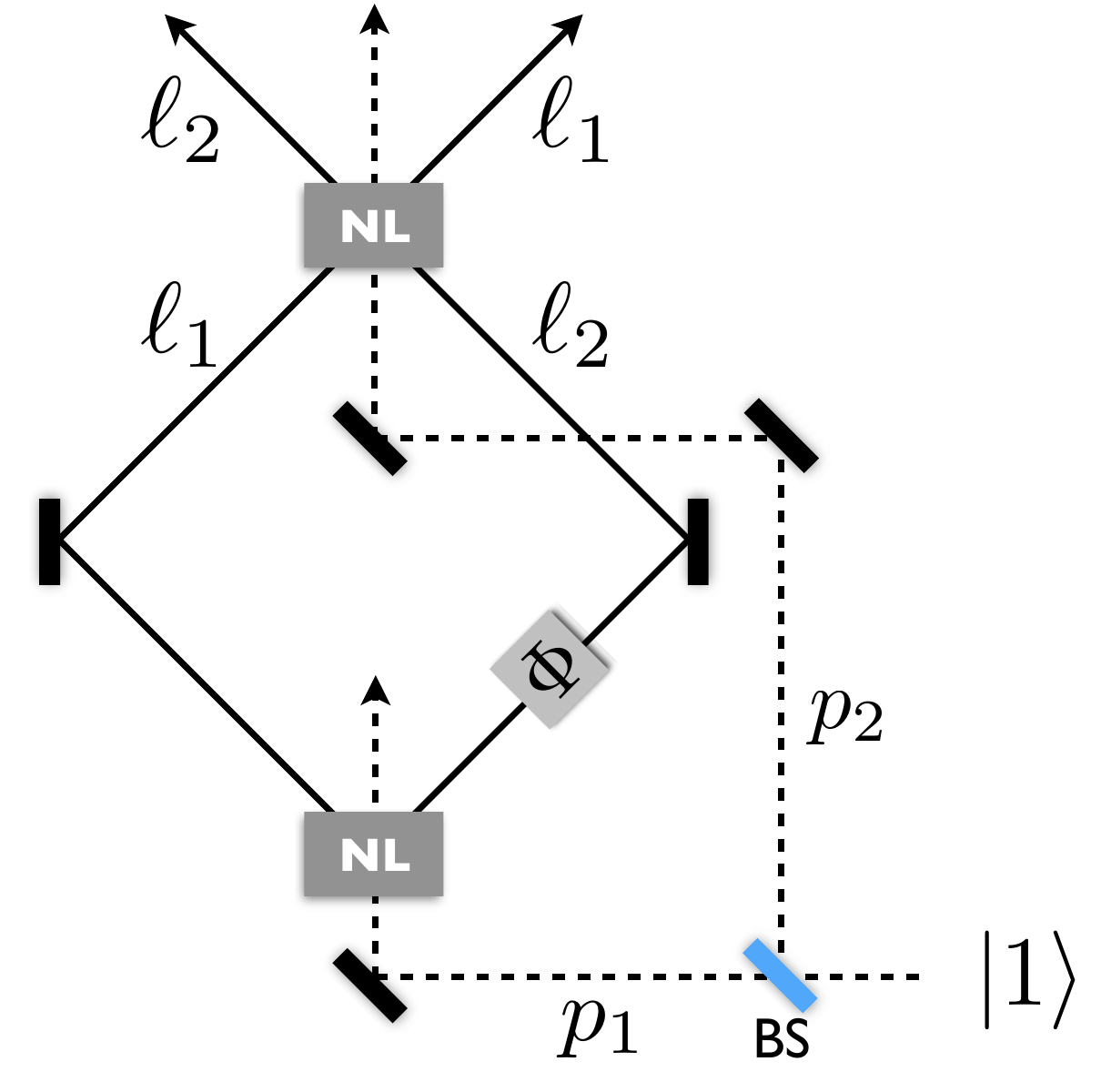}
%---------------------------------------------------------------------------------------
\hskip5mm
%---------------------------------------------------------------------------------------
\includegraphics[scale=0.37]{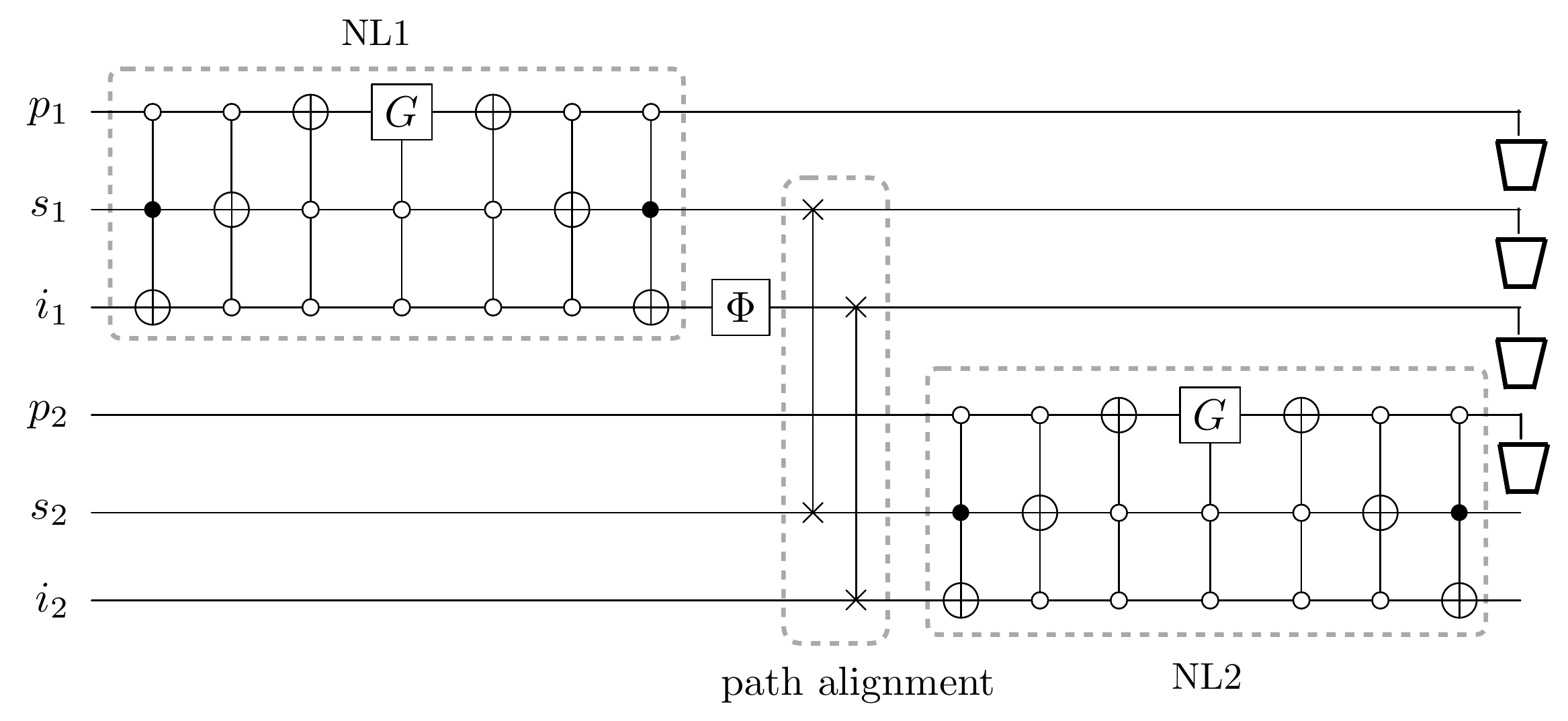}
\caption{Experimental setup (left) and quantum circuit (right) for the frustrated generation of photon pairs with a single-photon pump.}
\label{frustrated-mandel}
\end{figure}
%---------------------------------------------------------------------------------------

%%%%%%%%%%%%%%%%%%%%%%%%%%%%%%%%%%%%%%%%%%%%%%%%%%%%%%%%%%%%%%%%
\section{Appendix C: Details of the steps of the circuit for identification of a point-like object with undetected photons}
The steps of the circuit of Fig.~\ref{q-imaging-corr} after the application of the first NL, which yields $|\psi_1\rangle$ of Eq.~\eqref{psi-1}, are given in the following: 
\begin{align}
|\psi_2\rangle&:=\mathpzc{E}_O |\psi_1\rangle=|000\rangle_{s_1 i_1 w}-ig\alpha |H\rangle_{s_1}(T e^{i\gamma}|H 0\rangle+\sqrt{1-T^2}|0 H\rangle)_{i_1 w} |00\rangle_{s_2 i_2} \nonumber\\
|\psi_3\rangle&:=\textsc{swap}_{i_1 i_2}|\psi_2\rangle\nonumber\\
&~=|00\rangle_{s_1 i_1} |00\rangle_{s_2 i_2}|0\rangle_{w}-ig \alpha T e^{i\gamma}|H 0\rangle_{s_1i_1}|0 H\rangle_{s_2 i_2}|0\rangle_{w}-ig \alpha\sqrt{1-T^2}|H 0\rangle_{s_1 i_1} |00\rangle_{s_2 i_2} |H\rangle_w\nonumber\\
|\psi_4\rangle&:=U_{\textsc{nl2}}^{\alpha}|\psi_3\rangle\nonumber\\
&~=|\psi_0\rangle |0\rangle_{w}-ig\alpha (|0 H\rangle+T e^{i\gamma}|H 0\rangle)_{s_1 s_2}|0 H\rangle_{i_1 i_2}|0\rangle_{w}-ig \alpha\sqrt{1-T^2}|H 0\rangle_{s_1 s_2} |00\rangle_{i_1 i_2} |H\rangle_w.
\end{align}
The action of $\mathpzc{E}_O$, which is assumed not to affect the polarization of photons, is given in Refs.~\cite{Lemos,Ghalaii}. 
Tracing out over the idler path and path $w$, the state $|\psi_4\rangle\langle \psi_4|$ yields the same result expected from the experimental setup before the signal photons pass through the final beam splitter (Eq.~(22) of Ref.~\cite{Ghalaii}), 
\begin{align}
\Upsilon=&\frac{1}{2}\big[|H\rangle\langle H|\otimes |0\rangle\langle 0| + T e^{i\gamma} |H\rangle\langle 0| \otimes |0\rangle \langle H| + T e^{-i\gamma}|0\rangle \langle H|\otimes |H\rangle \langle 0| + |0\rangle\langle 0|\otimes |H\rangle\langle H|  \big]_{s_1 s_2}.
\end{align}

%%%%%%%%%%%%%%%%%%%%%%%%%%%%%%%%%%%%%%%%%%%%%%%%%%%%%%%%%%%%%%%%
\section{Appendix D: Extension of the matrix representation of NL in effective picture for higher number of modes}
We now show that the effective picture descriptions of the states and NLs lead to the reduction of tensor-product Hilbert space to a direct-sum Hilbert space. Consider a system with $2N$ paths, comprised of $N$ signal paths (labeled by $n_s \in \{1,\ldots,N\}$)  and $N$ idler paths (labeled by $n'_i \in \{1,\ldots,N\}$). 
We define the subspace $\mathpzc{H}_{n_s n'_i}=\mathrm{span}\{\ket{H H}, \ket{H V}, \ket{V H}, \ket{V V}\}_{n_s n'_i}$ attributed to signal path $n_s$ and idler path $n'_i$.

In a system comprising of coherently pumped NLs with weak laser beam, where the possibility of activation of two NLs at the same time is negligible, the total accessible Hilbert space is $\mathpzc{H}^{\mathrm{acc}}={\oplus}_{n_s,n'_i}\mathpzc{H}_{n_s n'_i}$. All state vectors of $\mathpzc{H}^{\mathrm{acc}}$ can now be represented as 
\begin{align}
|\psi^{\mathrm{eff}}\rangle=|\psi\rangle_{1_s 1_i}\oplus \cdots \oplus |\psi\rangle_{n_s n'_i} \oplus \cdots \oplus |\psi\rangle_{N_s N_i},
\end{align}
 in which $|\psi\rangle_{n_s n'_i}=P_{n_s n'_i} |\psi\rangle$, with $P_{n_s n'_i}=\sum_{k,l,p,q=H,V} |kl\rangle_{n_s n'_i} \langle pq |$ being the projector into the $\mathpzc{H}_{n_s n'_i}$.

Interestingly, the above decomposition also allows that the action of an NL on paths $n_s$ and  $n'_i$ be represented in a restricted fashion: $(\Lambda_{\textsc{nl}}^{\alpha})_{n_s n'_i}:~\mathpzc{H}_{n_s n'_i}\to \mathpzc{H}_{n_s n'_i}$; that is, $(\Lambda_{\textsc{nl}}^{\alpha})_{n_s n'_i}$ acts only on $\mathpzc{H}_{n_s n'_i}$. Thus an NL on paths $n_s$ and  $n'_i$ is given by $(\Lambda_{\textsc{nl}}^{\alpha})_{n_s n'_i}\oplus \openone_{\overline{n_s n'_i}}$ (see Fig.~\ref{direct-sum}). 
This implies that in this projective decomposition of the accessible Hilbert space NL acts nontrivially only on the vectors of the corresponding subspace.
%---------------------------------------------------------------------------------------
\begin{figure}[tp]
\includegraphics[scale=0.4]{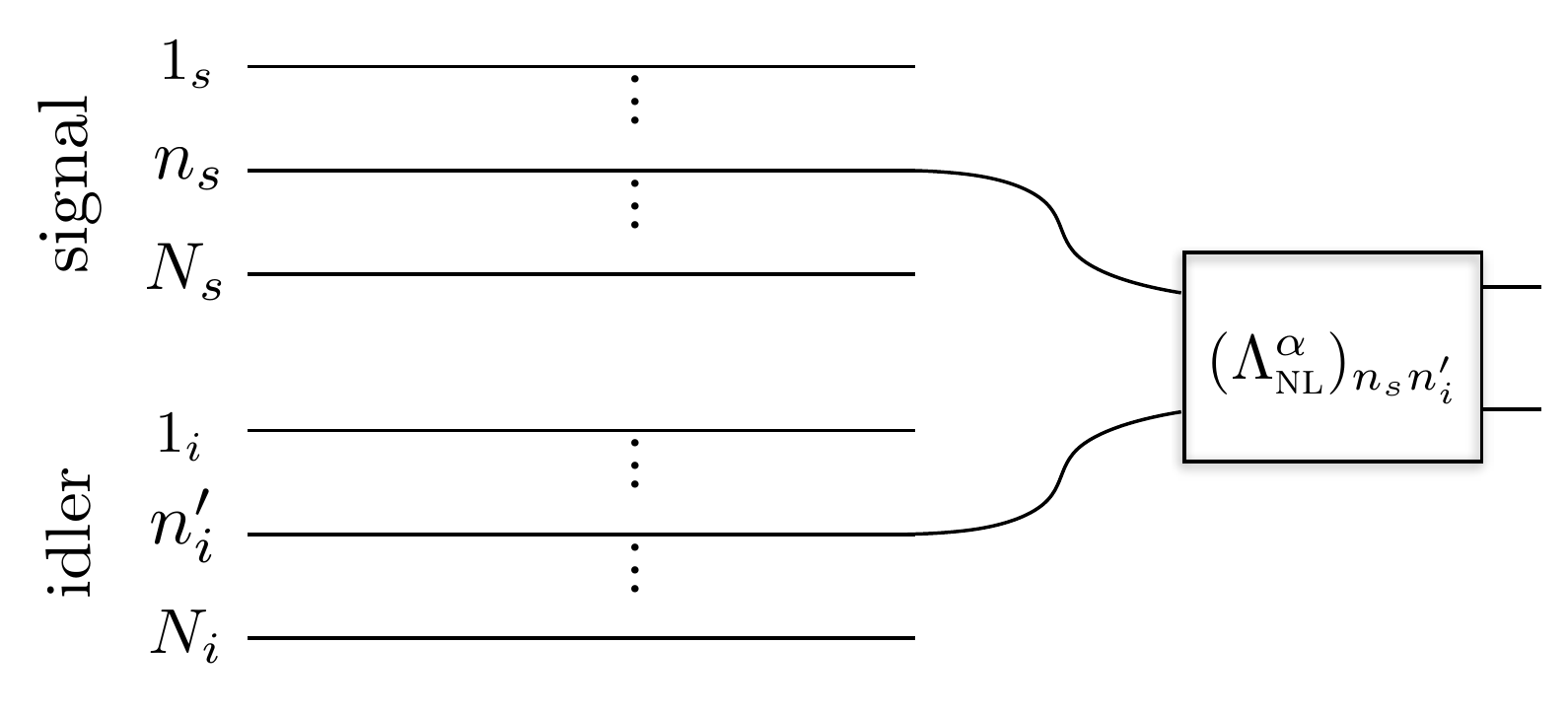}
\caption{A system with $N_s$ signal and $N_i$ idler paths, where an NL is applied on paths $n_s n'_i$}
\label{direct-sum}
\end{figure}
%---------------------------------------------------------------------------------------

%%%%%%%%%%%%%%%%%%%%%%%%%%%%%%%%%%%%%%%%%%%%%
\section{Appendix E: Details of the calculations of the modular superposition example}
Following the steps of Fig.~\ref{superposition} by using the matrix representation for NLs, one obtains that 
\begin{align}
\ket{00}&\overset{\Lambda^{\alpha}_{\textsc{nl}}}{\longrightarrow}\ket{H H}\overset{U^{(N)}}{\longrightarrow} U^{(N)} \ket{H H} \overset{\Lambda^{\alpha}_{\textsc{nl}}}{\longrightarrow} U^{(N)} \ket{H H}+ \ket{H H} \overset{U^{(N-1)}}{\longrightarrow} U^{(N-1)} U^{(N)} \ket{H H}+ U^{(N-1)} \ket{H H}  \nonumber\\
&~~~~~~~~~\cdots \nonumber\\
&\overset{U^{(2)}}{\longrightarrow} U^{(2)} \cdots U^{(N-1)} U^{(N)} \ket{H H}+U^{(2)} \cdots U^{(N-1)} \ket{H H}+ \cdots + U^{(2)} U^{(3)}\ket{H H}+U^{(2)}\ket{H H} \nonumber\\
&\overset{\Lambda^{\alpha}_{\textsc{nl}}}{\longrightarrow} U^{(2)} \cdots U^{(N-1)} U^{(N)} \ket{H H}+U^{(2)} \cdots U^{(N-1)} \ket{H H}+ \cdots + U^{(2)} U^{(3)}\ket{H H}+U^{(2)}\ket{H H}+\ket{H H} \nonumber\\
&\overset{U^{(1)}}{\longrightarrow} U^{(1)} \cdots U^{(N-1)} U^{(N)} \ket{H H}+ U^{(1)} \cdots U^{(N-1)}\ket{H H}+ \cdots 
+ U^{(1)} U^{(2)} \ket{H H}+ U^{(1)} \ket{H H} \nonumber\\
&~~~~~~~~\equiv \ket{\phi_N}+\ket{\phi_{N-1}}+\cdots+\ket{\phi_3}+\ket{\phi_2}+\ket{\phi_1}.
\end{align}

Alternatively, we can represent the whole action of the circuit in the matrix language as follows: 
%%%%%%%%%%%%%%
\begin{align}
& \left(\begin{smallmatrix}
1\\
0\\
0\\
0\\
0\\
\end{smallmatrix}\right)
\overset{\Lambda_{\textsc{nl}}^{\alpha}}{\longrightarrow}
 \left(\begin{smallmatrix}
1 & 0 & 0 & 0 & 0\\
1 & 1 & 0 & 0 & 0\\
0 & 0 & 1 & 0 & 0\\
0 & 0 & 0 & 1 & 0\\
0 & 0 & 0 & 0 & 1\\
\end{smallmatrix}\right)
\left(\begin{smallmatrix}
1\\
0\\
0\\
0\\
0\\
\end{smallmatrix}\right)=\left(
\begin{smallmatrix}
1\\
1\\
0\\
0\\
0\\
\end{smallmatrix}\right)\equiv\left(
\begin{smallmatrix}
1\\
\\
|HH\rangle\\
\\
\end{smallmatrix}\right)\nonumber\\
%%%%%%%%%%%%%%%%
&\overset{U^{(N)}}{\longrightarrow} 
\begin{pmatrix}
1 & \\
   & U^{(N)}
\end{pmatrix}
\left(
\begin{smallmatrix}
1\\
\\
|HH\rangle\\
\\
\end{smallmatrix}\right)=
\left(
\begin{smallmatrix}
1\\
\\
U^{(N)}|HH\rangle\\
\\
\end{smallmatrix}\right)
%%%%%%%%%%%%%%%%%%%
\overset{\Lambda_{\textsc{nl}}^{\alpha}}{\longrightarrow} 
\left(\begin{smallmatrix}
1 & 0 & 0 & 0 & 0\\
1 & 1 & 0 & 0 & 0\\
0 & 0 & 1 & 0 & 0\\
0 & 0 & 0 & 1 & 0\\
0 & 0 & 0 & 0 & 1\\
\end{smallmatrix}\right)
\left(
\begin{smallmatrix}
1\\
\\
U^{(N)}|HH\rangle\\
\\
\end{smallmatrix}\right)
=
\left(
\begin{smallmatrix}
1\\
\\
U^{(N)}|HH\rangle+|HH\rangle\\
\\
\end{smallmatrix}\right)
%%%%%%%%%%%%%%%%%
\overset{U^{(N-1)}}{\longrightarrow}\cdots \nonumber\\
%%%%%%%%%%%%%%%%%%%%%
& \overset{\Lambda_{\textsc{nl}}^{\alpha}}{\longrightarrow}
\left(
\begin{smallmatrix}
1\\
\\
U^{(1)}\cdots U^{(N)}|HH\rangle+U^{(1)}\cdots U^{(N-1)}|HH\rangle+\cdots +U^{(1)} U^{(2)}|HH\rangle +|HH\rangle\rangle\\
\\
\end{smallmatrix}\right).
\end{align}

%%%%%%%%%%%%%%%%%
%\end{widetext} 
\twocolumngrid
\end{widetext}
\end{document}